\documentclass[aps,prl,twocolumn,groupedaddress]{revtex4}

\usepackage{graphicx}

\begin{document}

\renewcommand{\vec}[1]{{\mathbf #1}}

\title{Enhanced mesoscopic correlations in dynamic speckle patterns}
\author{S.E. Skipetrov}
\email[]{Sergey.Skipetrov@grenoble.cnrs.fr}
\affiliation{Laboratoire de Physique et Mod\'elisation des Milieux Condens\'es/CNRS,\\
Maison des Magist\`{e}res, Universit\'{e} Joseph Fourier, 38042
Grenoble, France}

\date{\today}

\begin{abstract}
We study the temporal evolution of speckle patterns in transmission of short
wave pulses through a
disordered waveguide. In the diffuse regime,
the short-range spatial structure of speckles
is the same as for continuous wave (cw) illumination,
whereas the long-range correlation between distant speckle spots grows linearly with time
and can exceed its cw value. We discuss the physical
origin of this phenomenon, compare our results to recent microwave experiments, and suggest that
a similar linear growth with time should also be characteristic
for other mesoscopic interference phenomena.
\end{abstract}


\maketitle

In a disordered medium of size $L$, waves can either propagate ballistically (if $L \ll$
mean free path $\ell$), by diffusion (if $L \gg \ell$), or can be localized
(if $L$ exceeds the so-called localization length $\xi \geq \ell$)
\cite{anderson58,ping95,bart99,rossum99}.
The different regimes of wave propagation can be probed by sending into the medium
a short wave pulse and by studying the decay of the intensity $I(t)$ of waves
transmitted through the sample.
In the ballistic regime, $I(t)$ mimics the incident pulse, whereas in the diffuse regime
$I(t)$ exhibits appreciable sample-to-sample fluctuations with an ensemble average
$\left< I(t) \right>$ that decays exponentially with time
\cite{ping95,watson87}. The
characteristic time $t_D \sim L^2/D$ of this decay (the Thouless time)
features the diffusion constant $D$. In the localized regime,
the decay of  $\left< I(t) \right>$
slows down, suggesting that $D \rightarrow 0$
\cite{ping95,bart99}.
A similar situation is encountered when a brief voltage pulse is applied
to a disordered conductor (with $I(t)$ staying for the current
through the sample) \cite{alt87,muz95,mirlin00} or 
when a quantum particle is injected into an open chaotic cavity
(with $I(t)$ staying for the survival probability)
\cite{savin97}.
 
Recently, it has been realized that already in the diffuse regime the
closeness of the localization transition causes
the average intensity $\left< I(t) \right>$ to deviate from pure exponential decay
\cite{mirlin00,chabanov03a,skip04,cheung04}.
More precisely, the apparent diffusion constant decreases linearly with
time $t$ until $t$ becomes of the order of the Heisenberg time $t_H \gg t_D$.
Beyond the Heisenberg time, the so-called anomalously localized states dominate the dynamics
\cite{alt87,muz95,mirlin00}.
Thus, as time goes on, $\left< I(t) \right>$ behaves as if the system were undergoing
a continuous transition from the diffuse regime for
$t \sim t_D$
to the localized regime for $t > t_H$, when extended states have died out and only
localized states are relevant.
However, the intensity of waves in a disordered
medium fluctuates rapidly in space \cite{shapiro86}, giving rise to the so-called
speckle pattern, and hence the average value is not sufficient for its
statistical description.
The speckle pattern is observable by eye for light
(see, e.g., Ref.\ \cite{feng91})
and readily measurable for microwaves \cite{sebbah00} or ultrasound \cite{derode91}.
Does the speckle pattern $I(\vec{r},t)$ evolve with time and if it does, in
which way? Answering this question constitutes the principal subject of the present work.

The speckle pattern can be
quantitatively characterized by the correlation function of intensity fluctuations,
\begin{eqnarray}
C(\Delta\vec{r}, t) = \frac{ \left< \delta I(\vec{r}, t)
\delta I(\vec{r} + \Delta\vec{r}, t) \right> }{  \left< I(\vec{r}, t) \right>
\left< I(\vec{r} + \Delta\vec{r}, t) \right>}
\label{cdef}
\end{eqnarray}
where $\delta I(\vec{r}, t) = I(\vec{r}, t) - \left<  I(\vec{r}, t) \right>$.
Up to now, $C(\Delta\vec{r}, t)$ has only been studied for 
continuous-wave (cw) illumination in which case it is independent of $t$. $C$
can be split into a sum of short- and
long-range `mesoscopic' contributions \cite{feng88,feng91}.
The short-range contribution (usually denoted by $C_1$) is of order 1
for  $\Delta r <$ wavelength $\lambda$ but decays rapidly for $\Delta r > \lambda$,
identifying $\lambda$ as the size of an individual
speckle spot \cite{shapiro86,sebbah00}. Much attention has been given to
the long-range part of $C$ (denoted by $C_2$),
which survives for $\Delta r > \ell$ and can be interpreted as a correlation between
distant speckle spots \cite{stephen87,zyuzin87,pnini89,sebbah00}.
Detailed theoretical and experimental studies of $C_2$ have been undertaken for waves
in a quasi-1D disordered waveguide \cite{pnini89,sebbah00}, and we will also limit our consideration to this case.
In the diffuse regime (dimensionless conductance of the waveguide $g \gg 1$) and for cw illumination,
it has been shown that
$C_2(\Delta\vec{r}, t) = C_2^{\text{(cw)}}(\Delta\vec{r}) \sim 1/g \ll 1$
\cite{feng88,feng91,sebbah00}.
For pulsed illumination, the evolution of $C(\Delta\vec{r}, t)$ with time
has been recently studied in microwave experiments by Chabanov {\em et al.} \cite{chabanov04a}.
The experiments suggest that $C_1$ is independent of $t$ and that $C_2$ increases with time.
However, no theoretical model has been proposed so far to explain these observations.

In the present Letter we show that when a pulse of duration $t_p$ is sent into a disordered
waveguide, the short-range spatial structure of speckle pattern measured in transmission
through the waveguide is essentially the same as
for cw illumination (speckle spot size $\sim \lambda$).
In contrast, the long-range structure of
speckles appears to be seriously affected by the pulse nature of the incident wave:
We find that the long-range correlation of intensity fluctuations increases linearly with time,
\begin{eqnarray}
C_2(\Delta \vec{r}, t) &=&  C_2^{\text{(cw)}}(\Delta \vec{r})
\left( \alpha + \beta \frac{t}{t_D} \right)
\label{main}
\end{eqnarray}
and can exceed its cw value $C_2^{\text{(cw)}}(\Delta \vec{r})$.
Remarkably, the above result holds for \textit{any\/} pulse shape and duration,
provided that the correlation is measured at large enough times $t$ after the
extinction of the incident pulse and provided that
$t$ is well below the Heisenberg time $t_H$.
The parameters $\alpha$ and $\beta$ in Eq.\ (\ref{main}) depend on
the precise pulse shape, and (for a given pulse shape) on the ratio
$t_p/t_D$ of pulse duration to the Thouless time.
Besides, $\alpha$ exhibits a weak dependence on the optical thickness
$L/\ell$ of the waveguide,
the dependence becoming less and less pronounced as $L/\ell$ increases. 
Interestingly, we find that a similar linear growth with time should also be typical for other
types of mesoscopic correlations (e.g., the correlations of intensities corresponding to different
polarizations of scattered waves \cite{chabanov04b}) and fluctuations
(e.g., the fluctuations of the total transmission coefficient of a disordered
sample \cite{rossum99}), provided that they are measured for a pulsed incident wave.
Below we sketch the main steps of our analysis.

Let a pulse $\psi_0(t)$ be incident on the surface $z = 0$ of a quasi-1D disordered
waveguide of length $L \gg \ell \gg \lambda$ and with cross-section $A$. 
In the diffusion approximation, both the average intensity and
the short-range correlation function of intensity fluctuations for
the multiply scattered wave inside the waveguide can be found by summing ladder
diagrams, similarly to the cw case \cite{shapiro86}:
\begin{eqnarray}
&&\left< I(\vec{r}, t) \right> =
\frac{c}{4 \pi} \sum\limits_{n=1}^{\infty} \phi_n(z) \phi_n(\ell)
f^{(n)}(t, t)
\label{avint}
\\
&&\left< \delta I (\vec{r}, t)   \delta I (\vec{r}_1, t_1) \right>_{\text{short}} =
\left[ \frac{c}{4 \pi} \frac{\sin(k \Delta r)}{k \Delta r} e^{-\Delta r/2 \ell}
\right.
\nonumber \\
&&\left. \hspace{2cm} \times \sum\limits_{n=1}^{\infty} \phi_n(z) \phi_n(\ell)
f^{(n)}(t, t_1)
\right]^2
\label{c1}
\end{eqnarray}
where
$\phi_n(z) = (2/L)^{1/2} \sin(n \pi z/L)$,
$\Delta \vec{r} = \vec{r} - \vec{r_1}$,  $\Delta t = t - t_1 > 0$,
and we introduced
\begin{eqnarray}
f^{(n)}(t, t_1) = \int\limits_{-\infty}^{t_1} \mathrm{d} t^{\prime}\;
e^{-n^2 (t_1 - t^{\prime})/t_D} \psi_0(t^{\prime}) \psi_0^*(t^{\prime} + \Delta t)
\label{fn}
\end{eqnarray}
In what follows, we consider two different pulse shapes: a rectangular pulse 
$\left| \psi_0(t) \right|^2 = (1/t_p) \Pi(2 t/t_p)$,
where $\Pi(x) = 1$ if $\left| x \right| < 1$ and
$\Pi(x) = 0$ otherwise, and a Gaussian pulse
$\left| \psi_0(t) \right|^2 = (2/\sqrt{\pi} t_p)
\exp(-4 t^2/t_p^2)$, both of width $t_p$ and centered at $t = 0$.
For these two pulse shapes,
Eq.\ (\ref{fn}) can be evaluated analytically, but we omit the exact expressions for they
are rather cumbersome. In the present paper, we are mainly concerned with
`long times' that we define by requiring $t \gg t_p, t_D$ for
the rectangular pulse and $t \gg t_p^2/t_D, t_D$ for the Gaussian pulse.
At such long times, the $n = 1$ terms give the principal
contributions to Eqs.\ (\ref{avint}) and (\ref{c1}).
We find
\begin{eqnarray}
f^{(1)}(t, t_1) \simeq \frac{2 t_D}{t_p} \exp\left( -\frac{T}{t_D} \right) \sinh \left( \frac{
t_p - \left| \Delta t \right|}{2 t_D} \right)
\label{fnR}
\end{eqnarray}
if $\left| \Delta t \right| < t_p$ and $f^{(1)}(t, t_1) \equiv 0$ otherwise
for the rectangular pulse, and
\begin{eqnarray}
f^{(1)}(t, t_1) \simeq 
\exp\left( -\frac{T}{t_D} + \frac{t_p^2}{16 t_D^2} - \frac{\Delta t^2}{t_p^2} \right)
\label{fnG}
\end{eqnarray}
for the Gaussian pulse (we introduced $T = (t + t_1)/2$).

As follows from Eqs.\ (\ref{avint}), (\ref{fnR}) and (\ref{fnG}),
the average intensity decays exponentially with time, consistent with previous studies
\cite{ping95,watson87}.
The short-range correlation function of intensity fluctuations (\ref{c1})
decays rapidly to zero for $\Delta r > \lambda$, just as for cw illumination.
Hence, the finite duration of the incident pulse does not affect the short-range spatial
structure of the speckle pattern.
The correlation time $t_c$ of intensity fluctuations is set by $t_p$ for the Gaussian pulse,
whatever its duration,
whereas  $t_c \sim \min(t_p, t_D)$ for the rectangular pulse. 
This difference between long rectangular and Gaussian pulses is
due to the intrinsic difference in their shapes:
a long Gaussian pulse has slow variations of intensity, while
the intensity of a long rectangular pulse still jumps abruptly at $t = \pm t_p/2$.

Let us now consider the correlation between distant speckle spots, i.e. the long-range correlation
of intensity fluctuations. This can be calculated either by joining four
time-dependent ladder propagators using a Hikami box, in much the same way
as it is done for continuous
wave illumination \cite{stephen87}, or by generalizing the
steady-state Langevin approach \cite{zyuzin87,pnini89}. Both methods yield the same result:
\begin{eqnarray}
&&\left< \delta I (\vec{r}, t)   \delta I (\vec{r}_1, t_1) \right>_{\text{long}}
= \frac{2 \pi \ell c^2}{3 k^2} \int_V \mathrm{d}^3 \vec{r}^{\prime}
\int\limits_{-\infty}^{t} \mathrm{d} t^{\prime}
\int\limits_{-\infty}^{t_1} \mathrm{d} t_1^{\prime}
\nonumber \\
&& \hspace{2cm} \vec{\nabla}^{\prime} G(\vec{r}, t; \vec{r}^{\prime}, t^{\prime}) \cdot
\vec{\nabla}^{\prime} G(\vec{r}_1, t_1; \vec{r}^{\prime}, t_1^{\prime})
\nonumber \\
&& \hspace{2cm} \times \left< \delta I (\vec{r}^{\prime}, t^{\prime})
\delta I (\vec{r}^{\prime}, t_1^{\prime}) \right>_{\text{short}}
\label{didi}
\end{eqnarray}
where $G(\vec{r}, t; \vec{r}^{\prime}, t^{\prime})$ is the Green's function of the
diffusion equation and the spatial integral is over the volume $V$ of the sample.
In the rest of the paper, we restrict ourselves to $t = t_1$.
Substituting Eq.\ (\ref{c1}) into Eq.\ (\ref{didi}) we end up with
\begin{eqnarray}
\left< \delta I (\vec{r}, t)   \delta I (\vec{r}_1, t) \right>_{\text{long}}
&=& \frac{2}{3 g} \left( \frac{c}{4 \pi} \right)^2
\nonumber \\
&\times& \sum\limits_{m,n,p,q = 1}^{\infty} S_{mnpq} \;  T_{mnpq}
\label{didi2}
\end{eqnarray}
where
\begin{eqnarray}
S_{mnpq} &=& \frac{3 L}{\pi^4}
\phi_m(L-\ell) \phi_n(L-\ell) \phi_p(\ell) \phi_q(\ell)
\nonumber \\
&\times& \int\limits_0^L \mathrm{d} z \; \phi_m^{\prime}(z)
\phi_n^{\prime}(z) \phi_p(z) \phi_q(z)
\label{smnpq}
\\
T_{mnpq} &=& t_D^{-2}
\int\limits_{-\infty}^{t} \mathrm{d} t^{\prime}
\int\limits_{-\infty}^{t} \mathrm{d} t_1^{\prime} \;
e^{-m^2 (t - t^{\prime})/t_D - n^2 (t - t_1^{\prime})/t_D}
\nonumber \\
&\times& f_p(t^{\prime}, t_1^{\prime}) f_q(t^{\prime}, t_1^{\prime})
\label{tmnpq}
\end{eqnarray}

\begin{figure}
\includegraphics[height=6.3cm,angle=0]{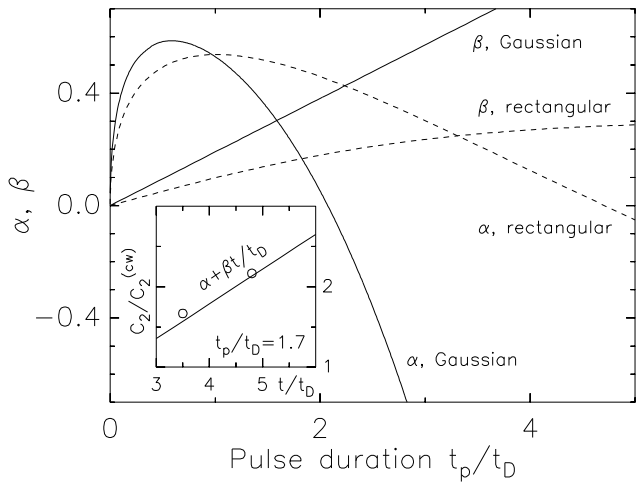}
\caption{\label{fig1}
Parameters $\alpha$ and $\beta$ that determine the time dependence of the long-range
intensity correlation function of intensity fluctuations 
$C_2(\Delta\vec{r}, t) = C_2^{\text{(cw)}}(\Delta\vec{r}) (\alpha + \beta t/t_D)$ in transmission of
rectangular (dashed lines) or Gaussian (solid lines) wave pulses of duration
$t_p$ through a quasi-1D disordered waveguide
($C_2^{\text{(cw)}}$ is the value of correlation for continuous (cw) illumination,
$t_D$ is the Thouless time, and the limit $L/\ell \rightarrow \infty$ is taken).
The inset shows the fit to the experimental data of Ref.\ \cite{chabanov04a} (symbols)
obtained by fixing $L/\ell = 10$, $L/L_a = 3$ and using $t_p/t_D$ as a fitting
parameter.}
\end{figure}

At times exceeding $t_D$ after turning on
of a very long ($t_p \gg t_D$) rectangular pulse, but before its extinction,
Eqs. (\ref{didi2})--(\ref{tmnpq}) yield the cw result $C_2^{\text{(cw)}} = 2/3 g$.
At long times, we can keep only the leading
terms in Eqs.\ (\ref{didi2}) [i.e., terms decreasing as $\exp(-2t/t_D)$]
and (\ref{avint}) [terms decreasing as $\exp(-t/t_D)]$.
This immediately gives Eq.\ (\ref{main}) with
\begin{eqnarray}
\beta &=& \frac{3}{2 \pi^2} \frac{\sinh(t_p/t_D) - t_p/t_D}{\sinh^2(t_p/2 t_D)}
\label{betar}
\end{eqnarray}
for the rectangular pulse and
\begin{eqnarray}
\beta &=& \frac{3}{(2 \pi)^{3/2}} \frac{t_p}{t_D}
\label{betag}
\end{eqnarray}
for the Gaussian pulse.
The derivation of the above analytic expressions is made possible by the fact that only one term
of Eq.\ (\ref{didi2})
(the term corresponding to $m = n = p = q = 1$) contributes to $\beta$.
In contrast, the calculation of $\alpha$ requires the summation of infinite
series in Eq.\ (\ref{didi2}).
We have carried out this summation numerically for $L/\ell \rightarrow \infty$ and show the
results in Fig.\ \ref{fig1} together with the analytical formulas (\ref{betar})
and (\ref{betag}) for $\beta$.

\begin{table*}
\caption{\label{tbl1}Summary of asymptotic results for parameters $\alpha$ and
$\beta$ governing the ratio of the time-dependent long-range correlation function
of intensity fluctuations
$C_2(\Delta\vec{r}, t)$  to its value $C_2^{\text{(cw)}}(\Delta\vec{r})$
for continuous wave (cw) illumination:
$C_2/C_2^{\text{(cw)}} = \alpha + \beta t/t_D$
($t_D$ is the Thouless time and $t_p$ is the duration of the
incident pulse).}
\begin{ruledtabular}
\begin{tabular}{ccccc}
 &\multicolumn{2}{c}{Rectangular pulse}&\multicolumn{2}{c}{Gaussian pulse}\\
  &Short pulse ($t_p \ll t_D$)&Long pulse ($t_p \gg t_D$)
&Short pulse&Long pulse\\ \hline
 $\alpha$&$\propto \sqrt{\frac{t_p}{t_D}}$&$0.67-\frac{3}{2 \pi^2} \frac{t_p}{t_D}$
&$\propto \sqrt{\frac{t_p}{t_D}}$&$-0.2 \left( \frac{t_p}{2 t_D} \right)^3$ \\
 $\beta$&$\frac{1}{\pi^2} \frac{t_p}{t_D}$
 &$\frac{3}{\pi^2}$&$\frac{3}{(2 \pi)^{3/2}} \frac{t_p}{t_D}$&
$\frac{3}{(2 \pi)^{3/2}} \frac{t_p}{t_D}$\\
\end{tabular}
\end{ruledtabular}
\end{table*}

Figure \ref{fig1} suggests that two different regimes can be realized depending
on the pulse duration. For short pulses ($t_p \ll t_D$), we find
$\alpha \propto (t_p/t_D)^{1/2}$ and $\beta \propto t_p/t_D$ for both pulse
shapes (see Table\ \ref{tbl1}), and the long-range correlation exceeds
its cw value if $t > t_D^2/t_p$.
In this regime, the incident pulse is too short for its shape to be resolved
in transmission, and hence we anticipate that apart from numerical factors,
our short-pulse results are fairly universal
(i.e., they hold for any pulse shape).
A different regime is realized for pulses of long ($t_p \gg t_D$) and intermediate
($t_p \sim t_D$) duration:
The long-range correlation function
exceeds its cw value through the whole range of validity of our analysis and
appears to be sensitive to the precise pulse shape (see Fig.\ \ref{fig1} and Table\ \ref{tbl1}).
Interestingly, for arbitrary relation between $t_p$ and $t_D$
the parameter $\beta$ seems to follow
a `universal' scaling relation $\beta \propto t_c/t_D$.

The growth of long-range correlation of intensity fluctuations with time
for Gaussian pulses of width $t_p \sim t_D$
has been clearly observed in experiments of Chabanov {\em et al.}
(see Fig.\ 3 of Ref.\ \cite{chabanov04a}),
thus providing an experimental support of our result (\ref{main}).
Unfortunately, the two available data points are insufficient to deduce the functional
dependence of $C_2$, either on $t$ or on $t_p$.
Nevertheless, the data can be fit with our theory
(see the inset of Fig.\ \ref{fig1}), provided that absorption
is incorporated in the theoretical analysis.
Absorption introduces an additional factor $\exp[-(t - t^{\prime})/t_a]$
in the expression for the intensity Green's function (here $t_a$ is the
absorption time). As a consequence, $m^2$ in Eq.\ (\ref{tmnpq}) for $T_{mnpq}$  
is replaced by $m^2 + (L/\pi L_a)^2$ (and similarly for $n^2$, $p^2$, and $q^2$)
with $L_a = (c t_a \ell/3)^{1/2}$
the absorption length.
By fixing $L/\ell = 10$ and $L/L_a = 3$ \cite{sebbah00,chabanov04a}
we obtain a satisfactory fit to the data with a reasonable value of $t_p/t_D = 1.7$.

The rise of the long-range correlation function $C_2$ with time can
be understood physically by taking advantage of the path picture of wave propagation in
disordered media.
In this picture, $C_2$ is due to the crossing of two wave paths inside
the sample \cite{akker04}. For a path of length $s$,
the probability of crossing is given by the ratio of the volume of a curvilinear tube of length $s$
and diameter $\sim \lambda$ to the total volume $V = AL$ of the waveguide,
yielding
$C_2 \sim s \lambda^2 / V$. For cw illumination, the typical path length $s \sim L^2/\ell$
yields $C_2^{\text{(cw)}} \sim 1/g$. For a pulsed source, $s = c t$ and
we obtain $C_2 \sim t/t_H$, where $t_H = g t_D$.
This simple qualitative reasoning is in complete agreement with the more rigorous
Eq.\ (\ref{main}) since the latter can be readily rewritten as
\begin{eqnarray}
C_2(\Delta \vec{r}, t) &=&  \alpha C_2^{\text{(cw)}}(\Delta \vec{r})
+ \frac{2}{3}\beta \frac{t}{t_H}
\label{main2}
\end{eqnarray}
The vanishing of $\alpha$ and $\beta$ (and hence of $C_2$) in the short pulse limit
$t_p/t_D \rightarrow 0$ (see Fig.\ \ref{fig1} and Table \ref{tbl1}) can be understood
by noting that in order to contribute to $C_2$, the two crossing wave paths should arrive at
the crossing point within a time
interval $\lesssim t_p$. Obviously, this becomes progressively less probable as $t_p$
decreases below the time of wave diffusion through the disordered sample $t_D$, leading to
vanishing of $\alpha$, $\beta$, and, consequently, of $C_2$ as $t_p/t_D$ tends to zero.

Equation (\ref{main2}) is expected to hold in higher-dimensional (2D, 3D) 
disordered systems as well,
with $\alpha$ and $\beta$ depending not only on $t_p/t_D$ but also
on the dimensionality and the experimental geometry.
It follows from Eq.\ (\ref{main2}) that the increase of long-range correlation
in speckle patterns with time favors bounded samples because otherwise
$t_H \propto V \rightarrow \infty$ and $t/t_H \rightarrow 0$.

In conclusion, we have shown that when a wave pulse is sent into a disordered waveguide,
the speckle pattern formed in transmission exhibits an enhanced long-range spatial correlation,
whereas the short-range correlation appears
to be identical to that for continuous wave (cw) illumination.
At long times, the long-range correlation grows linearly with time, whatever the
shape and the duration of the incident pulse, and can exceed its cw value.
Our analysis breaks down beyond the Heisenberg time, where the speckle
pattern is dominated by anomalously
localized states and the diffusion picture of wave propagation is not valid any more.
It is worthwhile to note that other types of mesoscopic phenomena should exhibit
a similar linear enhancement with time, when observed for a pulsed source of waves.
In particular, correlations between
waves of different polarizations and full probability distribution functions
of transmitted intensities have been recently measured in dynamic
microwave experiments \cite{chabanov04a}. Our analysis suggests that
the theoretical description of these measurements at times much shorter than the
Heisenberg time can be obtained by simply replacing
the dimensionless conductance $g$ by $g/(\alpha + \beta t/t_D)$ in formulas corresponding
to the stationary (cw) regime.

I am grateful to B.A. van Tiggelen and R. Maynard for numerous fruitful discussions
and for careful reading of the manuscript.


\end{document}